\documentclass{epl}
\usepackage{latexsym}

\title{Quantum decay of an open chaotic system:\\ a semiclassical approach}
\shorttitle{Quantum decay of open chaotic systems}
\author{Mathias Puhlmann \and Holger Schanz \and Tsampikos Kottos \and Theo
  Geisel}
\shortauthor{Mathias Puhlmann \etal}
\institute{
Max-Planck-Institut f\"ur Str\"omungsforschung
und Institut f{\"u}r Nichtlineare Dynamik\mbox{,} Universit{\"a}t G{\"o}ttingen,
Bunsenstra{\ss}e 10, D-37073 G\"ottingen, Germany
}

\def\/{\over}
\def\<{\langle}
\def\>{\rangle}
\def\({\left(}
\def\){\right)}
\def\[{\left[}
\def\]{\right]}
\def\L{{\cal L}}
\def\G{{\cal G}}
\def\i{{\rm i}}
\def\d{{\rm d}}
\def\e{{\rm e}}
\def\s{{S}}
\renewcommand{\nocite}[1]{}
\begin{document}

\maketitle

\begin{abstract}
  We study the quantum probability to survive in an open chaotic system in the
  framework of the van Vleck-Gutzwiller propagator and present the first such
  calculation that accounts for quantum interference effects. Specifically we
  calculate quantum deviations from the classical decay after the break time
  $t^{*}$ for both broken and preserved time-reversal symmetry.  The source of
  these corrections is identified in interfering pairs of correlated classical
  trajectories. In our approach the quantized chaotic system is modelled by a
  quatum graph.
\end{abstract}

A fundamental source of physical information are time-resolved decay
measurements in open quantum mechanical systems. While the radioactive decay
is a prominent paradigm, more recent experiments studied atoms in optically
generated lattices and billiards \cite{RSN97a,W+97a,MFR99,F+01b,K+01a,M+01c},
the ionization of molecular Rydberg states \cite{B+91a} and excitaton
relaxation in semiconductor quantum dots and wires \cite{B+99d,K+98}.  Most of
these examples, and also the complementary theoretical investigations of
quantum decay \cite{Dit00,JSS91,B+00c,WKB02,O+01a,
  B+01b,CMS97,Fra97,SS97a,CGM99, CMS99}, address the semiclassical regime of
systems with chaotic classical limit.

However, despite this broad interest there is no satisfactory semiclassical
theory for the observed quantum dynamics. It is known from numerical studies
and random-matrix theory (RMT) calculations \cite{CMS97,Fra97,SS97a,CGM99}
that the quantum survival probability $P(t)$ follows the exponential classical
decay $P_{\rm cl}(t)$ only up to a break time $t^{*}$.  This break time scales
with the number of open decay channels $L$ and the classical life time $t_{\rm
  cl}$ as $t^{*}\sim\sqrt{L}\,t_{\rm cl}$ \cite{CMS97}. For $t>t^{*}$ the
quantum decay law is a universal function which depends only on $L$, $t_{\rm
  cl}$ and the Heisenberg time $t_{H}$ and is qualitatively different from
$P_{\rm cl}(t)$ \cite{Fra97,SS97a,CGM99}. Up to now, none of these results was
accessible by semiclassical calculations and thus their applicability to
individual chaotic systems remained a matter of speculation.

\begin{figure}
\onefigure{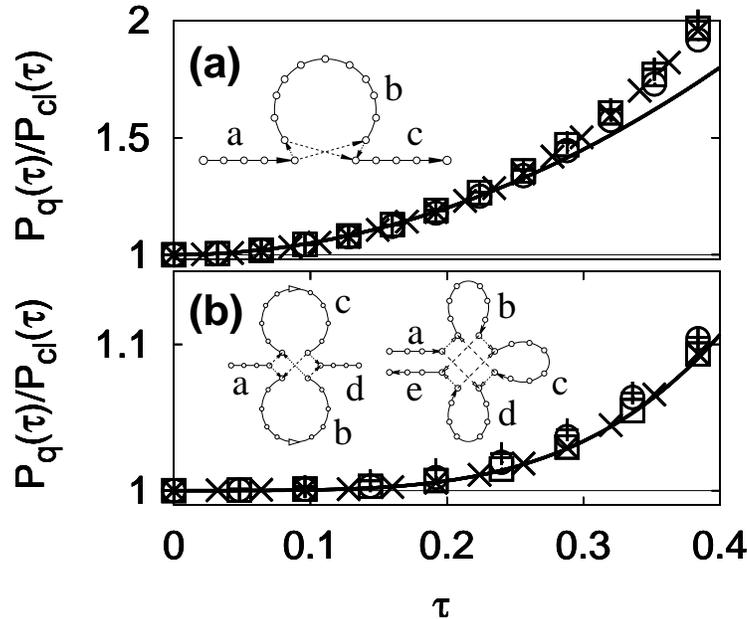}
\caption{
   The ratio between quantum and classical survival probability has been
   computed for a quantum graph with $L=10$ attached decay channels and
   $G=500 ({+})$, $750 ({\times})$, $1000 ({\Box})$, $2000
   (\circ)$ internal states. It is shown as a function of the scaled time
   $\tau=t/G$. Quantum corrections are visible as deviations of the data from
   the horizontal line. For small $\tau$ they follow (a) Eq.~(\ref{trs}) in the
   case with time-reversal symmetry and (b) Eq.~(\ref{ntr}) without. These
   semiclassical predictions account for the interference within the pairs of
   classical trajectories shown schematically in the insets. The two
   trajectories forming a pair are identical along the segments $a,b,\dots$ but
   differ in the crossing regions (solid vs. dashed arrows).}
\label{data}
\end{figure}

In this letter we show that a systematic semiclassical expansion for the
quantum decay can be based on the van Vleck-Gutzwiller propagator.
Specifically we obtain with this approach the above mentioned features
for the quantum probability to survive inside a quantized network (quantum
graph), which is one of the standard models in quantum chaos
\cite{KS97,KS00}. We identify the source of quantum deviations
from the classical decay in the interference between certain pairs of
correlated classical trajectories (inset of Fig.~\ref{data}). We have
calculated the quantum corrections analytically to leading order in time for
both, broken and preserved time-reversal symmetry (TRS). The resulting
expressions, Eqs.~(\ref{trs}), (\ref{ntr}) below, are in convincing agreement
with the corresponding numerical data (Fig.~\ref{data}).

Although we study in the present paper only systems with universal behaviour,
the semiclassical approach which we develop bears the potential to include
system-specific properties as well. It is natural that this important
advantage over any ad hoc random-matrix assumption comes at the price that
{\em semiclassical calculations cannot be completely independent of the
  underlying model}. Therefore some technical details of our calculation are
specific to quantized networks. Nevertheless our results are of interest also
beyond this class of models because the same pairs of correlated trajectories
will give rise to quantum corrections also in other chaotic systems. This
expectation is based on the analogy to weak-localization corrections in
spectral and transport properties \cite{SR01,Sie02,RS02,BSW02,BSW03,M+04}. 

It must be stressed, however, that the quantum corrections which we are
describing here go beyond the known weak localization effects as their
presence is not restricted to systems with time-reversal symmetry. In previous
studies of pairs of interfering classical trajectories it was always found
that they have no net effect on two-point correlation functions if
time-reversal symmetry is broken.  This applies to both, open and closed
systems \cite{SR01,Sie02,RS02,BSW02,BSW03,M+04}. Corrections have been found
in the case of shot noise \cite{SPG03}, but this quantity is a higher-order
correlator and involves groups of four correlated trajectories. So besides the
new physical context in which we study the effect of classical action
correlations the novelty in the present work lies in the fact that they can
have a non-vanishing effect on two-point correlators even without
time-reversal symmetry.

The dynamics of a quantum system is governed by a time-evolution operator $U$
propagating an initial state $|\psi_{0}\>$ from time $0$ to $t$ as
\begin{equation}
|\psi_{t}\>=U^{t}\,|\psi_{0}\>\,.
\end{equation}
For a closed system $U$ is unitary and therefore the norm of the initial state
is preserved $||\psi_{t}||^2\equiv 1$. In contrast, for an open system $U$ is
sub-unitary.  In this case the norm of the state, i.e., the survival
probability
\begin{equation}
||\psi_{t}||^{2}
=\<\psi_{t}|\psi_{t}\>
=\<\psi_{0}|U^{t}{}^{\dagger}U^{t}|\psi_{0}\>\,,
\end{equation}
decays from unity to zero as a function of time. We describe the smooth part
of this decay. Superimposed fluctuations are removed by considering an average
$P(t)=\<||\psi_{t}||^2\>_{\psi_{0},k}$ over the initial state $|\psi_{0}\>$.
An additional average is over the energy $k$. Assuming that the dynamically
relevant energy window contains $G$ states ($G\to\infty$ in the semiclassical
regime) and choosing a discrete basis $|m\>$, the average over $|\psi_{0}\>$
leads to
\begin{equation}\label{sp}
P(t)=G^{-1}\big\<\mbox{tr}\,U^{t}U^{t}{}^{\dagger}\big\>_{k}=
G^{-1}\sum_{m,n}\big\<
|(U^{t})_{mn}|^2
\big\>_{k}\,.
\end{equation}
$U$ can be approximated semiclassically by the van~Vleck-Gutzwiller propagator
\cite{vVl28,Gut71}
\begin{equation}\label{vanvleck}
(U^{t})_{mn}=\sum_{p}{\cal A}_{p}\,\exp(\i\,R_{p}(m,n;t)/\hbar)\,,
\end{equation}
i.e., a matrix element describing the transition $n\to m$ has contributions from 
all classical trajectories $p$ leading from $m$ to $n$ in time $t$. $R_{p}$ is the
action of the trajectory and ${\cal A}_{p}$ a complex amplitude combining a
stability factor and an additional phase from the Maslov index of $p$. 
With Eq.~(\ref{vanvleck}) we obtain for the survival probability
\begin{equation}\label{dsum}
P(t)=G^{-1}\,\big\<\sum_{p,q}{\cal A}_{p}{\cal A}_{q}^{*}\,
\exp(\i\,[R_{p}-R_{q}]/\hbar)\big\>\,,
\end{equation}
where the summation is over all pairs of classical trajectories $p,q$ which start
and end at the same point and have not left the system up to time $t$.

The classical probability to remain inside the system is reproduced if
Eq.~(\ref{dsum}) is restricted to the ``diagonal'' terms $p=q$. Then $P(t)$ is
simply a sum over all possible trajectories $p$ with corresponding
probabilities $|{\cal A}_{p}|^2$. It decays exponentially as,
\begin{equation}\label{pcl}
P_{\rm cl}(t)\sim\exp(-t/t_{\rm cl})\,,
\end{equation}
provided that the ergodic time of the chaotic flow is much shorter than the decay
time $t_{\rm cl}$ that is obtained from the relative phase-space area of the
opening \cite{BB90}.

To justify the diagonal approximation, which was developed originally in the
context of the spectral two-point correlator \cite{Ber85}, one observes that
in the semiclassical limit $\hbar\to 0$ the exponential in Eq.~(\ref{dsum})
represents rapidly oscillating phases which cancel upon averaging unless
$R_{p,q}$ are correlated. This is certainly the case for $p=q$, but recent
work on action correlations
\cite{A+93,CPS98,SR01,BSW02,M+04,SPG03} has shown that other
pairs contribute as well. The correlations in the actions stem from
the fact that these trajectories are composed of long segments where they
follow each other closely. In short crossing regions the order and/or
orientation of the segments is modified to yield different trajectories $p\ne q$.
It is convenient to represent trajectories by a symbolic code in which letters
$a,b,\dots$ stand for a whole segment. The codes of $p$ and $q$ are then
different words composed of the same set of letters. The number of segments,
i.e., the length of the code word determines the relative importance of the
action correlations.

Here we apply this scheme to the survival probability. As the trajectories
$p,q$ must start and end at the same points, their symbolic codes start and
end with the same letter. We need then at least three segments to obtain
different trajectories
\begin{equation}\label{ab'c}
p=[abc]\qquad q=[a\hat bc]\,,
\end{equation}
where $\hat b$ denotes the time-reversal of $b$ (see inset of
Fig.~\ref{data}a). These are precisely the pairs which are responsible for the
weak localization contribution to the conductance \cite{SR01} and they are
present for TRS only. A contribution which is present for systems with and
without TRS involves at least a four-letter word
\begin{equation}\label{acbd}
p=[abcd]\qquad q=[acbd]\,,
\end{equation}
(left inset of Fig.~\ref{data}b). However, it will be shown below that the
contributions from all such orbit pairs cancel. Therefore, in order to capture
the leading-order quantum corrections for systems without TRS, we need to
consider trajectory pairs with five segments of the form
\begin{equation}\label{adcbe}
p=[abcde]\qquad q=[adcbe]
\end{equation}
(right inset of Fig.~\ref{data}b). All other permutations with $a,e$ fixed are
excluded since two segments can be combined into a single one such that a pair
of the form (\ref{acbd}) results.  

The following calculation will be based on the trajectory pairs
(\ref{ab'c})-(\ref{adcbe}). It is plausible that they give the leading-order
quantum corrections because the two partner trajectories deviate only in a
minimum number of permutation points. This assertion is supported by our final
result and also by related work on spectral correlations and transport
\cite{SR01,BSW02,M+04,SPG03}.

The next step is to perform the summation in Eq.~(\ref{dsum}).
For this purpose we need explicit
expressions for the amplitudes and the actions of the classical trajectories and
therefore we consider a specific model system.  We assume  a quantized
network as it is commonly used, e.g., in mesoscopic physics \cite{Kuc02} and
quantum chaos \cite{KS97,Tan01,BSW02}. The discrete time-evolution operator on
a network with $B$ directed bonds has the form of a product of two $B\times B$
matrices
\begin{equation}\label{bsm}
U(k)=\s\times\,D(k)\,.
\end{equation}
Here $D_{mn}(k)=\delta_{mn}\,\e^{\i kl_{m}}$ is a diagonal matrix containing
phase factors which describe the free propagation along the bonds of the
network. $l_{m}$ ($m=1,\dots, B$) denotes the bond lengths which are chosen
to be incommensurate in order to avoid non-generic degeneracies. 
The wavenumber $k$ will be used for averaging, 
$\<\cdot\>_{k}=\lim_{k\to\infty}\,k^{-1}\int_{0}^{k}\d k'\,(\cdot)$. 
In particular, we have
\begin{equation}\label{average}
\<\e^{\i k(l_{m}-l_{n})}\>_{k}=\delta_{mn}\,.
\end{equation}
The matrix $\s$ fixes the topology of the underlying graph and the
classical transition probabilities between its bonds, $P_{n\to
  m}=|\s_{mn}|^{2}$. $P_{n\to m}$ specifies a Markovian random walk on the
graph which is the classical analogue of Eq.~(\ref{bsm}).  We make the
simplifying assumption that all transitions have equal probability,
\begin{equation}\label{transprob}
|\s_{mn}|^{2}=B^{-1}\,.
\end{equation}
The phases of $\s$ are still to be determined.  We consider first a closed
system. Then $\s$ must be unitary. As a second condition we note that for
preserved time-reversal symmetry the time-evolution operator is a symmetric
matrix up to a unitary transformation \cite{Haake}.  This is satisfied if
$\s=\s^{\rm T}$, since then $D^{{1\/2}}\,U\,D^{-{1\/2}}$ is symmetric. These
two assumptions together with Eq.~(\ref{transprob}) lead to a natural choice
for the matrix elements, $\s^{(1)}_{mn}=B^{-1/2}\exp(2 \pi\i
mn/B)$ \cite{Tan01}. In order to break time-reversal invariance we have to
destroy the symmetry of $\s$ but at the same time preserve the unitarity and
Eq.~(\ref{transprob}). This is achieved by a simple transformation exchanging
neighboring rows of $\s$, namely $\s^{(2)}= \Lambda\s^{(2)}$ with
$\Lambda_{mn}=\delta_{m,n-(-1)^{n}}$.  We have checked numerically that,
according to spectral statistics, a closed graph with $\s^{(\rm 1,2)}$
is a generic model for quantum chaos in the presence or
absence of TRS, respectively (see also \cite{Tan01,BSW02}).

Finally we need to open the system. A standard way to do this is to restrict
the time evolution to a subset $\G$ of $G<B$ internal states by the projector
$\Pi^{(\G)}$ with non-zero matrix elements $\Pi^{(\G)}_{mn}=1$ for $m=n\le G$,
\begin{equation}\label{uopen}
\tilde U=\Pi^{(\G)}\,U\,\Pi^{(\G)}\,.
\end{equation}
Effectively, the remaining $L=B-G$ bonds are then perfectly absorbing and play
the role of attached decay channels (leads). Nevertheless the set of these
bonds $\L$ influences the dynamics on $\G$, e.g., via the identity
\begin{equation}\label{unitarity}
\sum_{m\in \G}\s_{mn}\s^{*}_{mn'}
=\delta_{nn'}-\sum_{m\in \L}\s_{mn}\s^{*}_{mn'}\qquad\forall n,n'
\end{equation}
expressing the unitarity $\s\s^{\dagger}=I$ of the closed system.

The Heisenberg time of a network model is given by the number of bonds,
$t_{\rm H}=B$ \cite{KS97}. As our final results scale with $t_{\rm H}$ it will
be convenient to represent them in terms of the scaled time $\tau\equiv t/B$.
Whenever $\tau$ is used, we imply the semiclassical limit $t_{\rm H}\to
\infty$. This limit is taken with $\tau$ and $L$ fixed and we will keep only
the leading order terms in $t_{\rm H}=B$.

Substituting Eq.~(\ref{bsm}) into Eq.~(\ref{vanvleck}) and expanding
$(U^{t})_{mn}$ one finds that a trajectory $p$ of length $t$ on the graph is
just a sequence of bonds $p_{0},\dots,p_{t}$ with $p_{0}=n$ and $p_{t}=m$.
Amplitude and phase are given by ${\cal
  A}_{p}=\s_{p_{t}p_{t-1}}\dots\s_{p_{1}p_{0}}$ and
$R_{p}/\hbar=k(l_{p_0}+\dots+l_{p_{t}})$, respectively \cite{KS97}.  A
considerable simplification results from the fact that due to
Eq.~(\ref{average}) only orbit pairs with equal total lengths survive.
Consequently the phase factor in Eq.~(\ref{dsum}) is absent.

For $p=q$ we obtain $P_{\rm cl}(t)=\sum_{p}|{\cal
  A}_{p}|^2=(G/B)^{t}=(1-L/B)^{t}$, i.e., $P_{\rm cl}(\tau)=\exp(-L\tau)$.
This is equivalent to Eq.~(\ref{pcl}) with $t_{\rm cl}=B/L$. For the summation
over $p$ we have used that according to Eq.~(\ref{transprob}) the total
probability of a trajectory of length $t$ is $B^{-t}$, and that there are
$G^{t+1}$ such trajectories as each bond $p_{i}$ is summed over the whole set
$\G$.

In trajectory pairs composed of $s$ segments, the total time $t$ is the sum of
the lengths of the segments and of a contribution from each of the $s-1$ crossing
points $t=s-1+t_{a}+t_{b}+\dots$ Besides this constraint, $t_{a},t_{b},\dots$
can take any value $\ge 1$. This yields a combinatorial factor ${t-s\choose
  s-1}\sim (B\tau)^{s-1}/(s-1)!$.  Within the segments, amplitudes pair to
classical probabilities, Eq.~(\ref{transprob}), as in the diagonal
contribution above. Only the bonds right at the crossing points between
segments must be treated separately as a phase difference between $p$
and $q$ occurs there. We denote the factor from all crossing points for the moment
by $\Phi_{pq}$ and proceed first to the summation over the inner bonds of the
segments. For a segment $i$ of length $t_{i}$, containing $t_{i}-1$ inner
bonds, this yields $G^{-1}(G/B)^{t_{i}}$. The prefactor $G^{-1}$ is absent
for the first and the last segment as the first or last bond is not
involved in phase factors and can be summed. Including the explicit factor
$G^{-1}$ form Eq.~(\ref{dsum}) we have thus $G^{1-s}(G/B)^{t-s+1}\to
G^{1-s}\exp(-L\tau)$. Further, $G^{1-s}$ cancels
$B^{s-1}$ in the combinatorial factor above to leading order in $B$.
All terms together finally yield
\begin{equation}\label{pq}
P_{pq}(\tau)={\tau^{s-1}/ (s-1)!}\,\Phi_{pq}\,P_{\rm cl}(\tau)\,,
\end{equation}
where the phases $\Phi_{pq}$ remain to be calculated. We have
\def\afi{\bar\alpha}
\def\bin{\beta}
\def\bfi{\bar\beta}
\def\cin{\gamma}
\def\cfi{\bar\gamma}
\def\din{\delta}
\def\dfi{\bar\delta}
\def\ein{\varepsilon}
\begin{eqnarray}\label{phi}
&
\Phi_{a\hat bc}=\sum
\s_{\bin\afi}\s^{*}_{\bfi\afi}
\s_{\cin\bfi}\s^{*}_{\cin\bin}
&
\nonumber
\\
&
\Phi_{acbd}=\sum
\s_{\bin\afi}\s_{\cin\afi}^{*}
\s_{\cin\bfi}\s_{\bin\cfi}^{*}
\s_{\din\cfi}\s_{\din\bfi}^{*}
&
\\
&
\Phi_{adcbe}=\sum
\s_{\bin\afi}\s_{\din\afi}^{*}
\s_{\cin\bfi}\s_{\cin\dfi}^{*}
\s_{\din\cfi}\s_{\bin\cfi}^{*}
\s_{\ein\dfi}\s_{\ein\bfi}^{*}
&
\nonumber
\end{eqnarray}
where $\bar\alpha,\beta,\bar\beta,\dots$ stand for the last bond in $a$, the
first in $b$, the last in $b$, \dots and the summation for each index extends 
over the open graph $\G$ up to some crucial restrictions which we discuss now.
Consider a pair in the form (\ref{ab'c}) with $b=\bin b'\bfi$, assume
$\bin=\bfi$ and define $a'=a\beta$ and $c'=\bin c$. Then the very same orbit
pair can also be represented as $p=[a'b'c']$, $q=[a'\hat{b'}c']$. We exclude
this ambiguity by imposing the restriction $\bin\ne\bfi$. Similarly, for the
pairs (\ref{acbd}) we impose $\bin\ne\cin$, $\bfi\ne\cfi$ and for the pairs
(\ref{adcbe}) $\bin\ne\din$, $\bfi\ne\dfi$. With the help of these
restrictions and Eq.~(\ref{unitarity}) we have
\begin{eqnarray}
\Phi_{a\hat bc}&=&
\sum_{\afi\cin\in\L}\sum_{\bin\ne\bfi\in \G}
\s_{\bin\afi}\s^{*}_{\bfi\afi}
\s_{\cin\bfi}\s^{*}_{\cin\bin}
\\&=&
\sum_{\afi\cin\in\L}\(\sum_{\bin\bfi\in \G}
-
\sum_{\bin=\bfi\in\G}
\)\,
\s_{\bin\afi}\s^{*}_{\bfi\afi}
\s_{\cin\bfi}\s^{*}_{\cin\bin}\,.
\nonumber
\end{eqnarray}
In the second term with $\bin=\bfi$ the amplitudes $\s$ combine into
$B^{-2}$ and thus this term yields $L^{2}G/B^{2}$ which is negligible for
$B\to\infty$. In the first term we apply Eq.~(\ref{unitarity}) to perform the
(unrestricted) summation over $\bin,\bfi$. We obtain $LG^{2}/B^{2}\to L$
($B\to\infty$) plus another negligible term. Thus from Eq.~(\ref{pq}) with $s=3$
we obtain
\begin{equation}\label{trs}
P_{a\hat bc}(\tau)=(L/2)\,\tau^2\,P_{\rm cl}(\tau)\,.
\end{equation}
Applying repeatedly Eq.~(\ref{unitarity}) and estimating the order of the
resulting terms for $B\to\infty$ we find further
\begin{eqnarray}\label{ntr}
P_{acbd}(\tau)&=&0\,,
\nonumber\\
P_{adcbe}(\tau)&=&(L^2/24)\,\tau^4\,P_{\rm cl}(\tau)\,.
\end{eqnarray}
These expressions agree with the RMT \cite{Fra97,SS97a} to leading
order in $\tau$. This can be attributed to the rapidly mixing dynamics of our
model, see Eq.~(\ref{transprob}).  It is expected that higher orders are
reproduced as well, if in addition to Eqs.~(\ref{ab'c})-(\ref{adcbe})
trajectory pairs with more segments are included. Their omission is also
responsible for the remaining deviations between Eqs.~(\ref{trs}), (\ref{ntr})
and our numerical data (Fig.~\ref{data}). These deviations are larger for TRS,
where additional pairs with $s=4,5$ exist, while further corrections
independent of TRS have at least $s\ge 6$.

In conclusion, we have obtained semiclassical expressions for the
leading-order quantum corrections which determine the dynamics of the survival
probability after the break time $t^*$. As a corollary we are also able to
calculate the break time itself within the semiclassical theory: Using
the definition $c=P_{\rm q}(t^{*})/P_{\rm cl}(t^{*})$ ($c$ is some constant
threshold for relevant deviations from $P_{\rm cl}$) and substituting $P_{\rm
  q}=P_{\rm cl} +P_{a\hat bc}$ for TRS and $P_{\rm q}=P_{\rm cl}+P_{adcbe}$
for broken TRS we get $t^{*}\sim B/\sqrt{L}=\sqrt{L} t_{\rm cl}$ which agrees
with the break times evaluated numerically in \cite{CMS97}.

\acknowledgments
T.~K. and T.~G.~acknowledge support by a Grant from the GIF,
the German-Israeli Foundation for Scientific Research and Development.


\end{document}